\documentclass[twocolumn]{aastex631}

\usepackage[super]{nth}
\usepackage{savesym}
\savesymbol{tablenum}
\usepackage[T1]{fontenc}
\usepackage{inputenc}
\usepackage[range-units=brackets,product-units=single,multi-part-units=single,range-phrase={-},separate-uncertainty,print-unity-mantissa=false]{siunitx}
\usepackage{multirow}
\usepackage{appendix}

\graphicspath{{./}{figures/}}

\accepted{October 20, 2024}

\submitjournal{ApJ}
\shorttitle{230\,GHz time variability of the AGN in NGC~5044}
\shortauthors{Schellenberger et al.}

\begin{document}

\title{Probing the high frequency variability of NGC~5044: the key to AGN feedback}

\correspondingauthor{Gerrit Schellenberger}
\email{gerrit.schellenberger@cfa.harvard.edu}

\author[0000-0002-4962-0740]{Gerrit Schellenberger}
\affiliation{Center for Astrophysics $|$ Harvard \& Smithsonian, 60 Garden St., Cambridge, MA 02138, USA}

\author[0000-0002-5671-6900]{Ewan O'Sullivan}
\affiliation{Center for Astrophysics $|$ Harvard \& Smithsonian, 60 Garden St., Cambridge, MA 02138, USA}

\author[0009-0003-9413-6901]{Laurence P. David}
\affiliation{Center for Astrophysics $|$ Harvard \& Smithsonian, 60 Garden St., Cambridge, MA 02138, USA}

\author[0009-0007-0318-2814]{Jan Vrtilek}
\affiliation{Center for Astrophysics $|$ Harvard \& Smithsonian, 60 Garden St., Cambridge, MA 02138, USA}

\author[0000-0001-5725-0359]{Charles Romero}
\affiliation{Center for Astrophysics $|$ Harvard \& Smithsonian, 60 Garden St., Cambridge, MA 02138, USA}

\author{Glen Petitpas}
\affiliation{Center for Astrophysics $|$ Harvard \& Smithsonian, 60 Garden St., Cambridge, MA 02138, USA}
\affiliation{Massachusetts Institute of Technology, Massachusetts Ave., Cambridge, MA 02139, USA}

\author[0000-0002-9478-1682]{William Forman}
\affiliation{Center for Astrophysics $|$ Harvard \& Smithsonian, 60 Garden St., Cambridge, MA 02138, USA}

\author[0000-0002-1634-9886]{Simona Giacintucci}
\affiliation{Naval Research Laboratory, 4555 Overlook Avenue SW, Code 7213, Washington, DC 20375, USA}

\author[0000-0003-0685-3621]{Mark Gurwell}
\affiliation{Center for Astrophysics $|$ Harvard \& Smithsonian, 60 Garden St., Cambridge, MA 02138, USA}

\author[0000-0003-2206-4243]{Christine Jones}
\affiliation{Center for Astrophysics $|$ Harvard \& Smithsonian, 60 Garden St., Cambridge, MA 02138, USA}

\author[0000-0001-7509-2972]{Kamlesh Rajpurohit}
\affiliation{Center for Astrophysics $|$ Harvard \& Smithsonian, 60 Garden St., Cambridge, MA 02138, USA}

\author[0000-0001-5338-4472]{Francesco Ubertosi}
\affiliation{Dipartimento di Fisica e Astronomia, Universitá di Bologna, via Gobetti 93/ 2, I-40129 Bologna, Italy}
\affiliation{INAF - Osservatorio di Astrofisica e Scienza dello Spazio, via Gobetti 101, I-40129 Bologna, Italy}

\author[0000-0002-8476-6307]{Tiziana Venturi}
\affiliation{INAF - Istituto di Radioastronomia, via Gobetti 101, I-40129 Bologna, Italy}

\begin{abstract}
The active galactic nucleus (AGN) feeding and feedback process in the centers of galaxy clusters and groups is still not well understood. 
NGC~5044 is the ideal system in which to study AGN feedback. It hosts the largest known reservoir of cold gas in any cool-core galaxy group, and features several past epochs of AGN feedback imprinted as cavities in the X-ray bright intragroup medium (IGrM), as well as parsec scale jets. 
We present Submillimeter Array (SMA), Karl G. Jansky Very Large Array (VLA), James Clerk Maxwell Telescope (JCMT), and Atacama Large Millimeter/submillimeter Array (ALMA)  high frequency observations of NGC~5044 to assess the time variability of the mm-waveband emission from the accretion disk, and quantify the Spectral Energy Distribution (SED) from the radio to sub-millimeter band. 
The SED is well described by advection dominated accretion flow (ADAF) model and self-absorbed jet emission from an aging plasma with $\tau \sim \SI{1}{kyr}$. 
We find a characteristic variability timescale of 150 days, which constrains the ADAF emission region to about $\SI{0.1}{pc}$, and the magnetic field to $\sim \SI{4.7}{mG}$ in the jets and  and $\SI{870}{G}$ in the accretion disk. 
A longer monitoring/sampling will allow to understand if the underlying process is truly periodic in nature.

\end{abstract}


\section{Introduction} \label{ch:intro}
X-ray observations of the centers of relaxed galaxy clusters and groups show that these systems contain large amounts of hot X-ray emitting gas that should be radiatively cooling on timescales less than the Hubble time (see e.g., \citealp{Fabian1984-kj,Voit2011-wi,Stern2019-ph}). The observed radiation from the X-ray emitting plasma implied cooling rates up to $\SI{1000}{M_\odot\,yr^{-1}}$, predicting levels of star formation far in excess of those observed.  In the absence of a plausible energy input, the primary uncertainty in the original cooling flow scenario was the ultimate fate of the cooling gas. 

Only at the turn of the millennium did the first hints of a robust solution to the problem begin to appear with the comparison of high angular resolution X-ray and radio observations (e.g., \citealp{Bohringer1993-iv,Churazov2000-uy}). 
These analyses showed that supermassive black holes (SMBH) at the centers of X-ray bright atmospheres were radiatively faint, but mechanically powerful. Major modifications completely overturned and revolutionized the view of the cluster ``cooling flow'' scenario as active galactic nucleus (AGN)-induced cavities and shocks were found to be nearly universal properties of X-ray cooling atmospheres in hot gas rich clusters and groups  (e.g., \citealp{McNamara2000-pl,Blanton2003-os,Fabian2003-te,Forman2007-ew,Randall2010-gc,Werner2018-xb}). 
Although AGNs are able to offset the bulk of radiative cooling, some material does cool and is able to power the SMBH as part of the feedback cycle. 

Galaxy groups, such as NGC~5044, offer a unique window on the feedback cycle triggered by the central AGN. The shorter cooling times in groups mean we can sometimes observe multiple cycles of AGN activity visible by their traces in the intragroup medium (IGrM). The impact of feedback by the AGN is stronger in galaxy groups due to their shallower gravitational potential with respect to clusters, and together with the enhanced X-ray line emission a fine-tuned balance is required in order to establish a feedback cycle. Indeed, the reduced gas fractions observed in groups suggest that AGN feedback may over-heat them, expelling a significant fraction of the intra-group medium to large radii (see \citealp{Eckert2021-mm} for a review).

NGC~5044 is the X-ray brightest galaxy group in the sky with a wealth of multifrequency data available, making it an ideal object for studying correlations between gas properties over a broad range of temperatures. 
H$\alpha$ filaments, ro-vibrational H$_2$ line emission, [CII] line emission, and CO emission show that some gas must be cooling out of the hot phase (\citealp{Kaneda2008-xs,David2014-jn,Werner2014-vw}). 
Atacama Large Millimeter/submillimeter Array (ALMA), Atacama Compact Array (ACA), and IRAM single dish observations of NGC~5044 showed it to have the largest known amount of molecular gas among cool core galaxy groups (e.g., \citealp{Schellenberger2020-vl}). 
\cite{Schellenberger2020-vl} report hints for time variability of the continuum flux at 230\,GHz, and two absorption features in the CO(2-1) spectra.
\cite{Schellenberger2020-ji} used the Very Long Baseline Array (VLBA) to observe the source at 5 and $\SI{8.6}{GHz}$, and discovered a core-jet structure. From the almost identical brightness of the two jets, the authors concluded that the jets are  aligned close to the plane of the sky (also confirmed by \citealp{Ubertosi2024-ik}).  
The SED of AGN in NGC~5044 in the radio to sub-mm regime shows a turn over at the low frequency end, probably caused by synchrotron self-absorption, and a rising spectrum at mm-wavelengths. The central radio continuum source in NGC~5044 has a flux density of $45$ to $\SI{50}{mJy}$  at $\SI{230}{GHz}$ with a negative spectral index\footnote{We define the spectral index $\alpha$ as $S_\nu \propto \nu^{-\alpha}$, where $S_\nu$ is the flux density at frequency $\nu$}. \cite{Schellenberger2020-ji} conclude that emission from an advection dominated accretion flow (ADAF) explains the observed spectrum. However, missing data in the 10 to 100\,GHz regime introduces large uncertainties on several model parameters.  

The ADAF mechanism is based on the idea that gas is transported toward the AGN by a flow, and heated locally through the viscosity of the gas. A large amount of this energy is transported inward through ions of the accreted gas, and the rest is transported to the electrons and radiated via synchrotron and inverse Compton emission (and bremsstrahlung at higher frequencies than observed here, e.g., \citealp{Mahadevan1997-rx,Narayan1998-ia,Yuan2014-td}). 
The angular momentum of the cold infalling material establishes an accretion disk, which mediates the accretion rate. A stochastic variability of the mass accretion rate on short timescales has been proposed as a mechanism to link the time delays of AGN feedback and the infall of cold material onto the AGN (\citealp{Pope2007-zl,Pavlovski2009-er}). However, this process is not yet well understood.
While ADAFs have been observed in some nearby galaxies and Sgr A$^\star$ (\citealp{Donea1999-iv, Falcke1999-oq, Yuan2002-zx}) the cooling-flow in group central galaxies denotes an even more interesting environment in which ADAFs might be able to link the cold gas flow to the onset of small jets that start a feedback cycle. 
These previous ADAF detections are for SMBH masses of $\SI{2.6e6}{M_\odot}$ (Sgr A$^\star$, \citealp{Melia2001-ri}), $\SI{e6}{M_\odot}$ (M81, \citealp{Falcke1999-oq}), $\SI{3.5e7}{M_\odot}$ (NGC4258, \citealp{Falcke1999-oq}), while NGC~5044 is expected to have a SMBH mass of at least $\SI{e8}{M_\odot}$ (\citealp{David2009-hn}) or even $\SI{1.8e9}{M_\odot}$ (\citealp{Diniz2017-xw}).

In this paper we test the time variability of the AGN in NGC~5044 at mm wavelengths, and combine it with a refined SED that allows us to draw conclusions on the jet emission process, and the accretion characteristics. 
In section \ref{ch:data} we describe the new and archival data products used in this paper. In section \ref{ch:results} we present the result on the lightcurve, periodogram and SED analysis, and discuss these in section \ref{ch:discussion}. We present our summary in section \ref{ch:summary}. 
We adopt the heliocentric systemic velocity of NGC~5044 of $\SI{2757}{km\,s^{-1}}$ and a luminosity distance of $\SI{31.2}{Mpc}$ (\citealp{Tonry2001-wi}) to be consistent with previous works (e.g., \citealp{David2017-ig,Schellenberger2020-ji,Schellenberger2020-vl}).
This results in a physical scale in the rest frame of NGC~5044 of $\SI{1}{\arcsec}=\SI{150}{pc}$. Uncertainties are given at the $1\sigma$ level throughout the paper.

\section{Observations}
\label{ch:data}
In the following, we describe the data reduction of our recent Submillimeter Array (SMA) data revealing interesting variability in the mm lightcurve, and the archival high frequency VLA and James Clerk Maxwell Telescope
 (JCMT) data reduction to improve constraints derived from the radio/mm SED. 

\subsection{SMA}\label{ch:data_sma}
Starting in 2021 January, we monitored the flux of NGC~5044 at 230\,GHz with the SMA, an interferometer on Mauna Kea in Hawaii with eight 6\,m diameter dishes (\citealp{Gurwell2007-wz,Grimes2020-ik,Grimes2024-zo}). 
By 2024 May thirty successful observations had been performed, with a typical monthly cadence (PI Schellenberger, see Tab. \ref{tab:SMA} for a list of projects and observations). However, from August until early December each year, NGC~5044 is not observable for the SMA due to the 25 degree solar avoidance zone, and the generally poorer phase stability in the afternoon (when NGC~5044 would be visible during this time of the year). 
Since NGC~5044 is a point source at 230\,GHz\footnote{The extent of the VLBA jets at 6.7\,GHz is about $\SI{20}{mas}$, and the smallest restoring beam of SMA at 230\,GHz is $\SI{0.5}{\arcsec}$. Therefore, it is safe to assume that NGC~5044 is a point source for SMA at mm wavelengths.}, we utilized all SMA configurations, and did not place constraints on the receiver setup other than the use of the 230/240 receivers, while the subband configuration can vary among the observations. 
Table \ref{tab:SMA} also lists the observing time, on-source time, number of active antennas, the used flux and bandpass calibrators, and the observing frequency range. 
We included frequent phase reference scans of \verb|1337-129| between the target scans. 

The data reduction was performed using the \textit{pyuvdata} package (\citealp{J-Hazelton2017-of}) to convert data from the SMA native format to measurement sets (MS), which are passed to the subsequent analysis with the \textit{CASA} package (version 6.5.4, \citealp{McMullin2007-ed,The-CASA-Team2022-mr}). 

The SMA SWARM correlator (\citealp{Primiani2016-dc}) achieves a $\SI{140}{kHz}$ channel resolution, and we bin by a factor of 64 to 256\, spectral channels per sideband per receiver and per each of the 6 subbands.
In a first step we decide on the bandpass and flux calibrator, with the latter, depending on elevation and integration time,  sometimes taken from the end of a previous observation in the same frequency setup. 
We manually flag the flux- and bandpass calibrators by plotting the amplitude versus time and channels, and we remove the edge channels (2.5\% at each edge). 
We set the fluxscale to 'Butler-JPL-Horizons 2012', which is the standard for solar system objects, including our flux calibrators, Ceres, Pallas, Vesta, and Titan. 
An initial phase calibration of the bandpass calibrator allows us to derive the bandpass, which was visually inspected. 
Using these solutions, we prepare phase calibrator solutions (phases per integration interval, and amplitudes per scan). We bootstrap the flux calibration including bandpass and phase calibration to all fields, and image the bandpass and phase calibrators for verification. 
We then image the target and apply a phase self-calibration if the noise level is  improved, which is almost always the case. 
We finally determine the source flux and uncertainty using the CASA task \verb|imfit|. 
We add a systematic uncertainty of 5\% to account for flux density scale uncertainties when comparing SMA fluxes with other instruments. 

\subsection{VLA}\label{ch:data_vla}
NGC~5044 was observed with the VLA in A configuration for absolute flux measurements at $\SI{1.3}{cm}$ and $\SI{0.7}{cm}$ (K and Q band, 22 and 44\,GHz, respectively). 
These observation are part of project 21B-149 (PI Schellenberger) and were performed on May 26, 2022, with 26 available antennas. Of the total observing time of 1 hour, 24 minutes were on-target (12 minutes each band). 
The source 3C286 was observed as amplitude and bandpass calibrator, and several short scans of J1337-1257 were phase reference calibration scans.  

The data analysis was done with the automated CASA VLA pipeline (\citealp{The-CASA-Team2022-mr}), using the \cite{Perley2017-ug} flux scale. The quality of the output product was confirmed, and logfiles were screened for errors. 
We \verb|split| the spectral windows of the K and Q band observations into separate measurement sets, and proceeded with two (one) cycles of phase self-calibration for K (Q) band. 
The final images were clean and the noise decreased significantly to $\SI{30}{\mu Jy\,bm^{-1}}$ and $\SI{0.5}{mJy\,bm^{-1}}$ for K and Q band, respectively. The restoring beams were $109 \times \SI{59}{mas}$ and $57 \times 
\SI{33}{mas}$ for K and Q band, respectively. 
We note that the largest resolvable scale at this configuration is 2.4\,arcsec for K, and 1.2\,arcsec for Q band. These are much larger than any possible extent of the source emission, which is expected to come from the mpc-scale accretion disk, or the VLBA jets which are expected to be faint at these frequencies (\citealp{Schellenberger2020-ji}). 
Our final fluxes are $\SI{17.3}{mJy}$ and $\SI{24.6}{mJy}$ at $\SI{22}{GHz}$ and $\SI{44}{GHz}$, respectively. 
We conservatively add a 5\% systematic error in quadrature, which dominates the total uncertainty of the fluxes. 

\subsection{JCMT}\label{ch:data_jcmt}
The James Clerk Maxwell Telescope on Mauna Kea, Hawaii, is a 15\,m  sub-mm observatory, with the continuum bolometer array instrument SCUBA-2 installed in the Cassegrain focus. 
The 15 observations of NGC~5044 since 2014 at $\SI{850}{\mu m}$ ($\SI{352.7}{GHz}$) and $\SI{450}{\mu m}$ ($\SI{666.2}{GHz}$) include 4 new SCUBA-2 observations (see Tab \ref{tab:jcmt}), which were not performed or publicly available at the time of publication of \cite{Schellenberger2020-ji}. 
We download the already processed and calibrated datasets 
(\citealp{Holland2013-rb,Chapin2013-hd,Mairs2021-zf}) from the Canadian Astronomy Data Centre (CADC, \citealp{Ball2011-yq}). Our analysis procedure follows \cite{Schellenberger2020-ji}. 
We determine the fluxes by fitting the individual images with 2D Gaussians in sherpa which is provided with the CIAO 4.16 package (\citealp{Fruscione2006-wt,Burke2023-qm}). 
For each of the two bands we perform simultaneous image fits, and link several parameters, such as the beam size and the amplitudes for observations on the same day. 
The fitted beam sizes are $\SI{11.0(1)}{arcsec}$ and  $\SI{8.5(2)}{arcsec}$, for 850 and $\SI{450}{\mu m}$, respectively, which is relatively consistent with \cite{Dempsey2013-ya}. 
We note that the most recent measurement from February 2, 2021, is almost contemporaneous with an SMA observation. 

For $\SI{850}{\mu m}$ the higher source flux and lower instrument noise allows us to derive individual fluxes for each observation with typical noise uncertainties between 3 and 4 mJy (see Tab. \ref{tab:jcmt} for details). For our SED of NGC~5044 we consider only the last measured value $\SI{850}{\mu m}$ in 2021 of $\SI{47.1(31)}{mJy}$. 
A reliable flux measurement is not possible for individual images at $\SI{450}{\mu m}$, so we constrain the (average) flux in a simultaneous fit of all observation, which gives $\SI{39.6(24)}{mJy}$. 
In order to have a self-consistent value with respect to the other fluxes in the SED, we use this measured flux to estimate the expected value in February 2021: we compute an average $\SI{850}{\mu m}$ flux using the same fitting method, which we combine with the 2021 measurement to derive a scaling factor ($\num{0.829}$) for the average $\SI{450}{\mu m}$ flux to obtain an expected 2021 flux at $\SI{450}{\mu m}$. 

\section{The central SMBH in NGC~5044}\label{ch:results}
\begin{figure*}[t]
    \centering
    \includegraphics[width=0.99\textwidth]{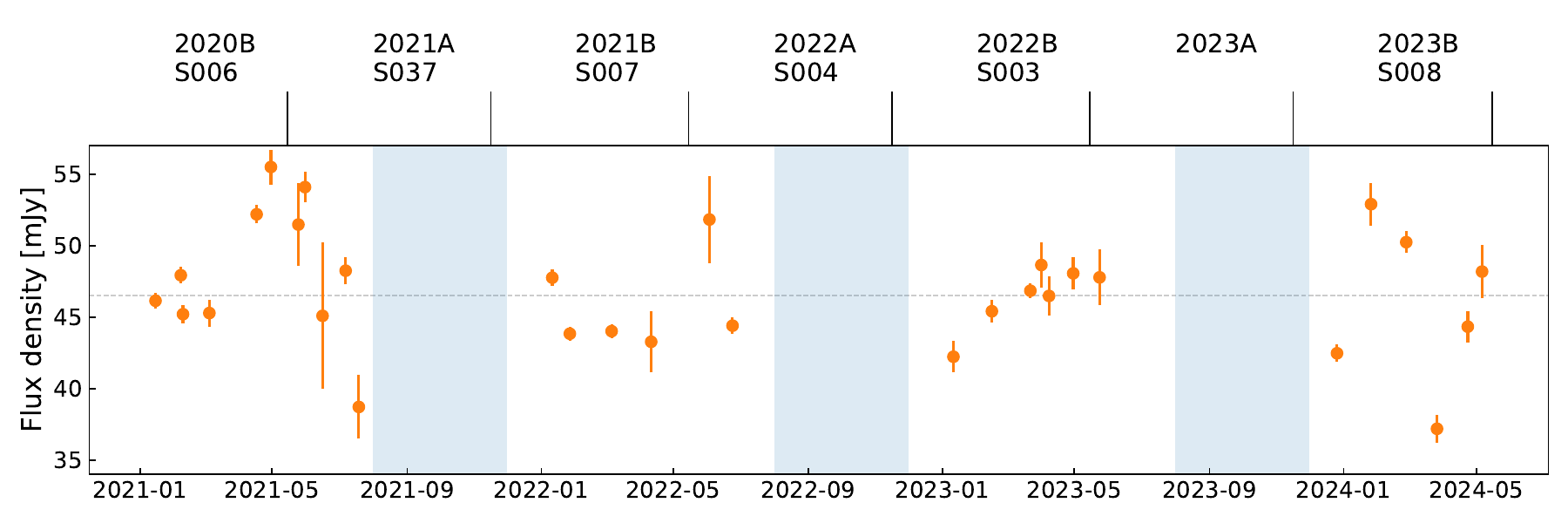}
    \caption{SMA Lightcurve at 230\,GHz. Blue shaded areas mark the unobservable part of each year, where the NGC~5044 is too close to the sun or nighttime observations are not possible.}
    \label{fig:SMA}
\end{figure*}
Our new data reveals intriguing results for the time variability and the spectral energy distribution (SED) of the central SMBH in NGC~5044. We first present our findings on the mm-lightcurve (section \ref{ch:lightcurve}) and the derived periodogram analysis (section \ref{ch:periodogram}), and link this for a time-consistent SED in section \ref{ch:sed}.  

\subsection{The lightcurve at mm-wavelengths}\label{ch:lightcurve}
\begin{figure}[t]
    \centering
    \includegraphics[width=0.49\textwidth]{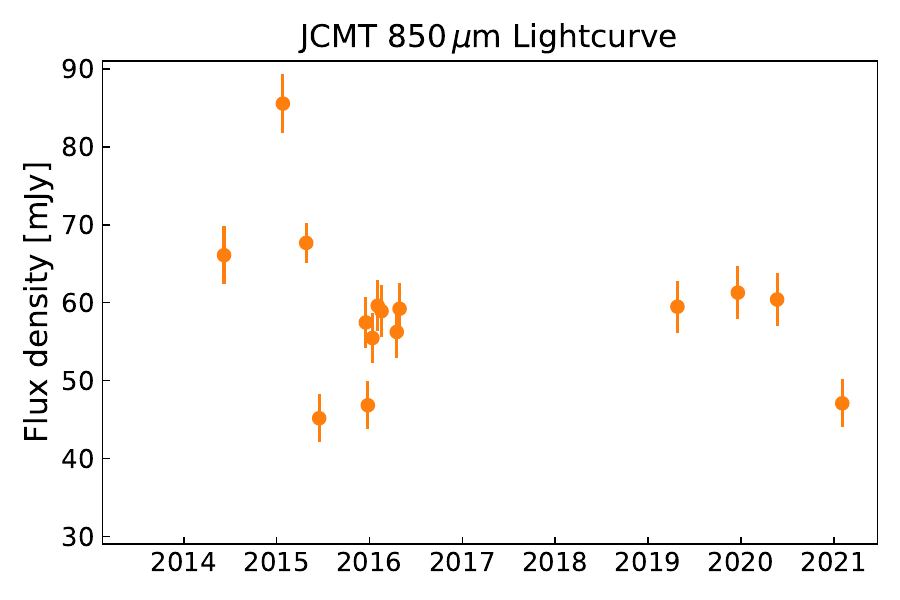}
    \caption{JCMT lightcurve at $\SI{850}{\mu m}$ (353\,GHz).}
    \label{fig:JCMT}
\end{figure}
A change in the mm continuum flux was first noticed in ALMA data taken in different epochs and configurations (\citealp{Schellenberger2020-vl}). 
Our new observations of 3.5 years of SMA monitoring observations of NGC~5044 allow us to analyze the mm variability at 230\,GHz (flux meausrements are listed in Tab. \ref{tab:smaflux}) which we present in Fig. \ref{fig:SMA}. 
The uncertainties of the individual flux measurements are typically between 0.5 and 1\,mJy, which was the requirement of the proposed observations. However, on several observing days the achieved noise was highly affected by weather/opacity, the unavailability of particular antennae owing to engineering work, and/or the standard flux calibrators being unavailable (see Tab. \ref{tab:SMA}). Observations on these days have larger uncertainties. 
Whenever observations were completely unusable they were usually repeated by the SMA staff in a timely manner. 
Soon after the start of our monitoring program in the first half of 2021 we find a peak in the lightcurve as the flux increases by 20\% from the baseline level of about 46\,mJy to about 55\,mJy within a few months. By mid 2021 the flux appears to be back to the baseline level, before NGC~5044 becomes unobservable. 
After the observations restart in 2022 we find decreasing fluxes from 48\,mJy to a consistent baseline until late June 2022 of 44\,mJy. 
The 2023 lightcurve does not include any strong peak, but contains measurements with $5-10\%$ lower fluxes than the previously assumed baseline. 
Finally, our findings for the first half of 2024 indicate a very disturbed lightcurve with variability on a 1 to 2 month scale. 
The SMA lightcurve covers the time of our VLA measurements (see section \ref{ch:data_vla}), and the ALMA 170\,GHz measurements (project 2021.1.00766.S, PI Rose), which allows us to rescale these other instruments to a common time.

We present the JCMT $\SI{850}{\mu m}$ lightcurve in Fig. \ref{fig:JCMT} with measurements between 2014 and 2021. 
The sampling is highly uneven, with, for example, 6 measurements within 5 months around 2016 and no observation in the following 3 years. 
The overall scatter is 22\%, and we see a slight downturn in 2021, where there is overlap with SMA data at 230\,GHz.

\begin{figure}[t]
    \centering
    \includegraphics[width=0.49\textwidth]{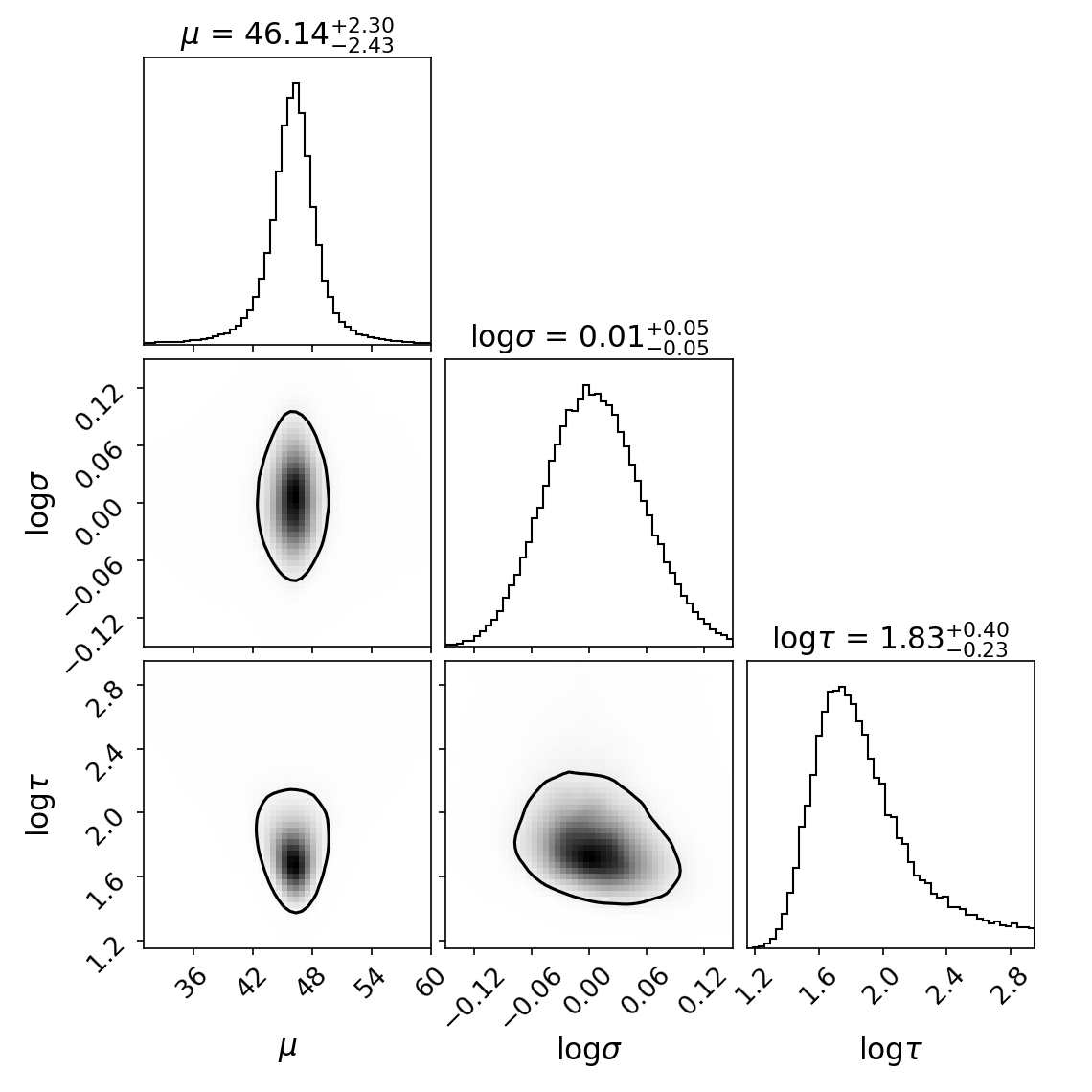}
    \caption{Corner plot illustration of the MCMC posterior parameters (see Eq. \ref{eq:drw}) from the DRW fit to the SMA lightcurve. The contours show the 68\% confidence level. Note that the mean flux $\mu$ and the variability $\sigma$ are measured in mJy, the timescale $\tau$ in days. }
    \label{fig:DRW}
\end{figure}
\subsection{Damped Random Walk}

The damped random walk model (DRW, \citealp{Kelly2009-mi}) is a stochastic process with an exponential covariance matrix, which has been successful in understanding non-periodic lightcurve data. DRW has been applied to optical and millimeter lightcurves of Sgr A$^\star$  (e.g., \citealp{Macquart2006-zb,Dexter2014-dz,Bower2015-bb,Wielgus2022-lk}), and quasars (e.g., \citealp{Kelly2009-mi,Kozlowski2009-vj,MacLeod2010-ly,Zu2013-ym,Chen2023-uz}). 
The DRW model assumes that changes in flux are related to a random walk motion with a damping component. The change in flux $dX(t)$ are described as
\begin{equation}\label{eq:drw}
    dX(t) = - \frac{1}{\tau}\left( X(t) - \mu \right)dt + \sigma dB(t)~,
\end{equation}
$X(t)$ denotes the flux at time $t$, $\mu$ is the mean value of the lightcurve, $B(t)$ is the Brownian motion, a non-stationary random walk process with a $1/f^2$ power spectrum, $\sigma$ is the short-term variability, and $\tau$ is the characteristic timescale for a random walk to return to the mean of the lightcurve. As pointed out by \cite{Kelly2009-mi} the first term on the right hand side of Eq. \ref{eq:drw} drives the flux back to the mean value (within a characteristic timescale $\tau$), while the second term creates random perturbations, e.g., caused by changes in the accretion or the magnetic field. The resulting power spectrum is flat (white noise) for frequencies below a threshold, and corresponds to red noise on frequencies above the threshold. 
We use the drw4e package\footnote{https://pypi.org/project/drw4e/} with the Gaussian error model to fit our lightcurve data. We set the prior distributions on $\mu$ to include the observed average range of fluxes, flat priors for $\sigma$ and $\tau$, before we run the Markov Chain Monte Carlo with $\num{100000}$ samples. Our results are presented in Fig. \ref{fig:DRW}. All parameters are reasonably well constrained and the best fit values and $1\sigma$ intervals are shown on top of each parameter's 1D distribution. 
The average flux $\mu$ and the short term variability $\sigma$ are both in agreement with expectations (mean and scatter of the lightcurve), while the timescale is $\tau = 67^{+102}_{-28}$ days. We note that the posterior distribution of $\tau$ has a significant tail towards higher values, mostly due to the sparse sampling.

\subsection{Periodogram}\label{ch:periodogram}
The DRW model tested successfully for a non-periodic timescale on the order of several months. However, we also present other methods to test for variability in lightcurves. 
A complication is posed by sparsely and unevenly sampled lightcurves which are set by the SMA observations, despite the efforts to have an approximate monthly cadence.
The shortest sampled timescales are on the order of a few days, corresponding to frequencies $\sim \SI{3e-1}{day^{-1}}$, while the longest time intervals are a few years, corresponding to frequencies $\sim \SI{e-3}{day^{-1}}$. 
We generally do not expect variability on timescales shorter than the light crossing time of the inner region of the accretion disk, which is a few Schwarzschild radii (corresponding to several days, depending on SMBH mass and truncation radius, see section \ref{ch:sed}).

\begin{figure}[t]
    \centering
    \includegraphics[width=0.49\textwidth]{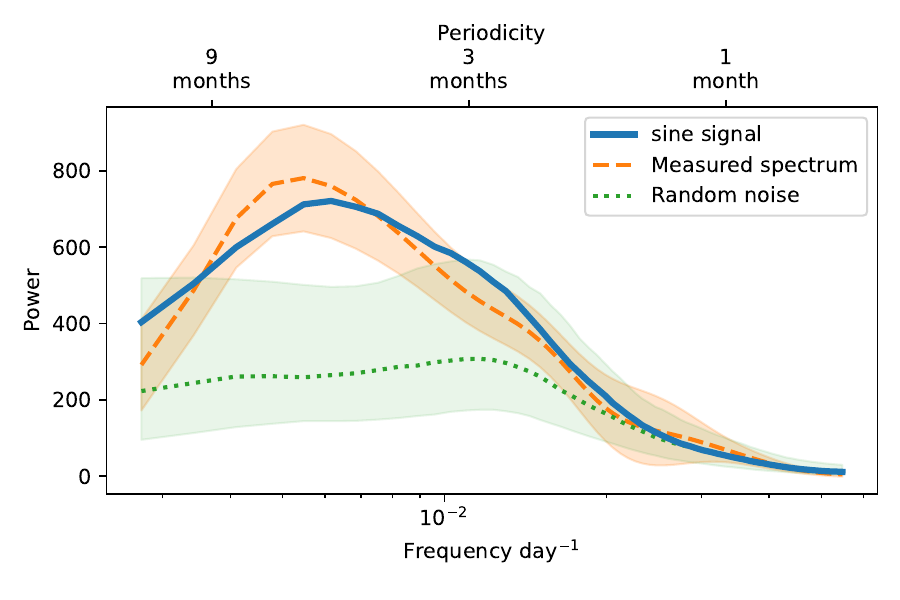}
    \caption{Power spectrum of the SMA lightcurve based on the Arevalo-method (orange), and from simulated data of random noise (green) and a sine with 136 day period (blue).}
    \label{fig:Arevalo_sin}
\end{figure}

\begin{figure}[t]
    \centering
    \includegraphics[width=0.49\textwidth]{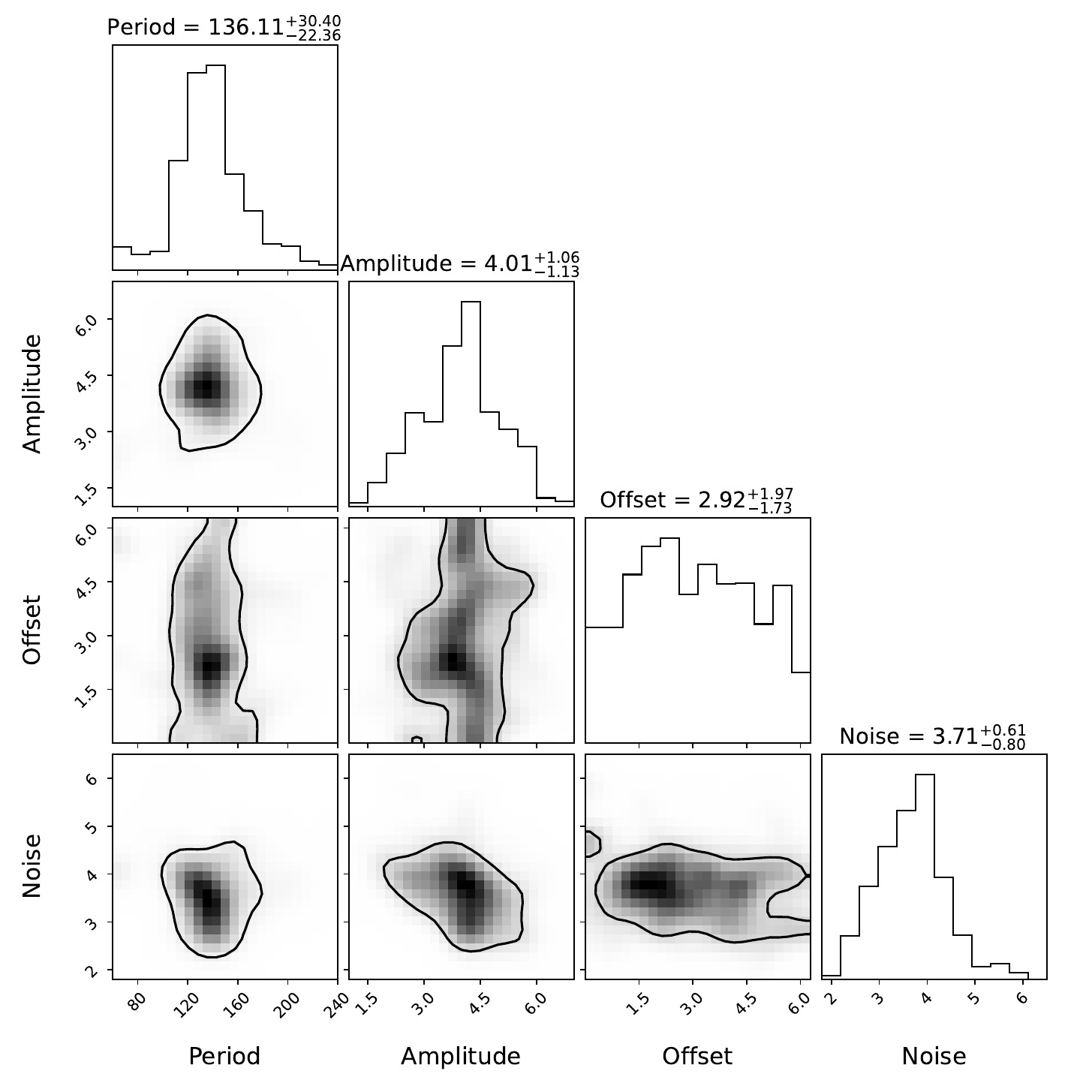}
    \caption{Illustration of the MCMC posterior parameters from the power spectrum fit of a simulated sine lighturve to the observed power spectrum, calculated via the \cite{Arevalo2012-kf} method. The contours show the 68\% confidence level. Note that units for the plotted quantities are days for the period, mJy for the sine amplitude, rad for the phase offset, and mJy for the Gaussian noise.}
    \label{fig:Arevalo_corner}
\end{figure}

In contrast to the DRW model, we use the $\Delta$-variance method by \cite{Arevalo2012-kf} to detect periodic signals throught the power spectrum of  sparsely sampled data. This method has mostly been applied to determine the power spectra of fluctuations in two-dimensional surface brightness images with arbitrary masking (\citealp{Zhuravleva2015-zs,Romero2023-uq,Romero2024-fz,Dupourque2024-zd}), but can be used for one-dimensional problems as well (as stated in \citealp{Arevalo2012-kf}). 
The concept stems from Parseval's Theorem, which effectively equates the integral of the power spectrum to the integral of the variance. In effect, the difference of Gaussians (Mexican Hat) acts as a not-so narrow $\delta$ function whereby one can specify a frequency at which one can recover the power.
The variance of this difference is proportional to the power of the frequency that was tested. 
We verified that potential biases of this method do not affect our sampling region (timescales between a month and a year). 
The orange dashed line in the left panels of Fig. \ref{fig:Arevalo_sin} shows the measured spectrum using this algorithm. The overall curve is very smooth, with a clear peak around 6 months, and some minor peak at 3 months. 
We use a sine input signal to derive a power spectrum with the same mask as the observations to derive a more precise periodicity. Our sine model has four free parameters, the period, the sine amplitude, a phase offset (ranging from 0 to $2\pi$, and the amplitude of a random Gaussian noise term. We run an MCMC with $\num{100000}$ samples to fit the power spectrum of the simulated signal to the measured power spectrum, both with the same observing mask. We note that the phase offset parameter should not change the power spectrum, and it remains an unconstrained parameter. Our posterior distributions are shown in Fig. \ref{fig:Arevalo_corner}. The period is well constrained with $136^{+30}_{-22}\,\si{d}$, and the amplitude of the sine is $\SI{4.0(11)}{mJy}$, which is slightly larger than amplitude the random noise distribution of $3.7^{+0.6}_{-0.8}\,\si{mJy}$. 
The best fit sine-power spectrum is over-plotted in Fig. \ref{fig:Arevalo_sin} in blue. We also show the noise contribution to the sine power spectrum in green (Fig. \ref{fig:Arevalo_sin}), which clearly dominates at higher frequencies, and is not flat as naively expected. 
Since we have a random test signal (green curve) which represents our null hypothesis, we can compare its statistics (e.g., $\chi^2$) with the best fit model's statistics using the F-test (\citealp{Protassov2002-jo}). 
We find that the p-value is $\num{3.3e-5}$, which makes the sine model more than $4\sigma$ significant with respect to a random noise model. This makes the hypothesis of the SMA lightcurve being random unlikely.

For completeness, we also show a Lomb-Scargle periodogram in appendix \ref{ch:ls}, which does not allow us to robustly distinguish between a variable signal and random noise. The Arevalo-method shows that the the lightcurve is most likely a periodic signal with 136 day period (about 5 months). 
We note that the timescales derived from the periodogram broadly agrees with the DRW model timescale. This model independent inference strengthens the conclusion of a variability timescale. 

\subsection{The Spectral Energy Distribution}\label{ch:sed}
A mm-upturn of the continuum fluxes in the central radio source in NGC~5044 was first noticed by \cite{David2014-jn,David2017-ig} when comparing ALMA 230\,GHz data with lower frequency radio fluxes. 
\cite{Schellenberger2020-ji} have measured and modeled the SED of the radio continuum source in NGC~5044 from $\sim \SI{200}{MHz}$ to $\SI{600}{GHz}$, and concluded that the emission appears to be well parametrized by a jet model with a self absorption component, combined with an advection dominated accretion flow (ADAF) model. The combination of a jet and ADAF model has been applied to other Low Luminosity AGNs (LLAGN) and low-ionization nuclear emission-line region (LINER) AGN (\citealp{Quataert1999-vx,Nemmen2014-ol,Wu2007-xz, Xie2016-fj,Yan2024-ay}), several of them in the vicinity of M87/Virgo.  

The role of dust emission from central galaxy in NGC~5044 can potentially create another obstacle, since its emission model starts to become important at wavelengths of $\lesssim \SI{1}{mm}$. Dust emission is often described by  a modified blackbody model (e.g., \citealp{Zielinski2024-hr, Vaillancourt2002-zv, Chuss2019-el}). \cite{Temi2018-rr} have measured the dust component in NGC~5044 based on Spitzer observations and derived a temperature of around $\SI{30}{K}$. Based on their model we can estimate the total flux from dust at $\SI{230}{GHz}$ to be $\SI{0.8(2)}{mJy}$. This is similar to our statistical uncertainties of the SMA flux measurements. \cite{Temi2018-rr} also show a dust map based on starlight extinction, which reveals a filamentary dust structure that extends $\SI{10}{\arcsec}$. Our SMA beam is generally much smaller $\SI{5}{\arcsec}$, which deminishes the dust contribution for our measurements even further.

Flux densities of the central radio source are listed in \cite{Schellenberger2020-ji} from Giant Metrewave Radio Telescope (GMRT), VLA, VLA Sky Survey (VLASS), ALMA and JCMT. 
However, the lack of reliable, recent measurements between 10 to $\SI{100}{GHz}$ added large uncertainties on some of the model parameters of the jet component. 
This gap is filled by our recent VLA K and Q band observations, and we also add more recent ALMA and JCMT data. 
The synergy of the mm-lightcurve and the SED is evident; 
our detailed mm-lightcurve allows us to ``rectify'' fluxes to a common epoch (Feb. 2021). The ``corrected'' SED can be fit with models to obtain parameters on feeding and feedback history, which in turn can be used to interpret the mm-variability in terms of the accretion history. 

Our combined model consists of 12 parameters, which we describe here. Our ADAF emission model follows the description in \cite{Schellenberger2020-ji}. 
The first parameter is the SMBH mass $M$, which was determined by \cite{David2009-hn} through the \cite{Gebhardt2000-pt} $M-\sigma$ scaling relation to be $\SI{2.3e8}{M_\odot}$. However, \cite{Diniz2017-xw} used a more recent relation (\citealp{Kormendy2013-hd}) and a precisely measured velocity dispersion and concluded a SMBH mass of $\SI{1.8e9}{M_\odot}$, which is over 7 times larger than the previous estimate used by \cite{Schellenberger2020-ji}. We leave $M$ free to vary in our fit.  
The accretion rate, $\dot M$, is also left free in our fit and is expected to be low $\dot M < \SI{0.1}{M_\odot\,yr^{-1}}$. It can be compared with star formation rates and cooling rates (e.g., \citealp{McDonald2018-xg}). Variability in the ADAF flux can most easily be explained by a changing accretion rate. 
The inner ADAF truncation radius, $r$, is set to $3R_S$, where $R_S=\frac{2G M}{c^2}$ is the Schwarzschild radius.  
The viscosity parameter, $\alpha$, is also fixed to $0.3$, which is typically used for ADAFs (\citealp{Liu2013-nq}). 
The gas-to-total pressure $\beta$ can also be referred to as the magnetic parameter, since the total pressure is the sum of the gas and magnetic pressure, which gives the identity $\frac{p_{\rm mag}}{p_{\rm gas}} = \frac{1}{\beta}-1$, where $p_{\rm mag} = \frac{B^2}{8 \pi}$. 
We typically use a fixed $\beta=0.5$, which means that the magnetic and gas pressure are equal, but also have a test-case where the magnetic pressure is negligible, $\beta = 0.99$. 
The last ADAF parameter in our model is $\delta$, which quantifies the heat energy distribution between electrons and ions. $\delta$ is expected to be close to the mass ratio of electrons and protons $\delta \approx \frac{1}{1836}$, but we leave it free to vary with a normal prior centered at $\frac{1}{1836}$ and a $\sigma_\delta = \num{2e-4}$.

For the jet emission model we use the \verb|synchrofit| package (\citealp{Turner2018-lq, Turner2018-qj}), which contains routines to calculate synchrotron emission of an aging plasma following the JP approach (\citealp{Jaffe1973-cs}), the KP formalism that does not include any electron pitch angle scattering (\citealp{Kardashev1962-re, Pacholczyk1971-ze}), and the Continuous Injection model (CI, \citealp{Komissarov1994-kv}). All three models have basic parameters, such as a normalization $J_0$ (left free to vary), an injection spectral index $\alpha_{\rm inj} = \frac{s-1}{2}$, where $s$ is the spectral index of the energy distribution of electrons injected into the plasma, $N(E)dE \propto E^{-s} dE$, and is fixed to $s=2$ (see e.g., \citealp{Carilli1991-qy}), and a break frequency $\nu_{\rm break}$. The CI model also has an additional parameter, $\kappa=\frac{T_{\rm off}}{\tau}$, that describes the fraction of time that the AGN is injecting energy into the plasma. Note that the JP and KP models do not have an active injection of ``fresh'' electrons, and describe the emission of an aging plasma which was injected at $t=0$. 
Lastly, the synchrotron self absorption (SSA) model (\citealp{Blandford1979-oy,Rybicki2008-dg}) contains only two parameters, the spectral index $\alpha_{\rm SSA}$ and the the break/turn-over frequency $\nu_{\rm SSA}$ between the optically thin and thick regime. Both SSA parameters have no priors. While for a perfectly optically thick plasma $\alpha_{\rm SSA}=-2.5$ is expected, shallower slopes have often been observed (e.g., \citealp{Laor2008-mu,Ishibashi2011-an,Kim2021-lp}). 

\begin{deluxetable*}{ccccccccccccc}
 \tablecaption{SED Fitting results.\label{tab:sed_result}}

\tablehead{\colhead{Run} & \colhead{Jet} & \colhead{$M$} & \colhead{$\dot M$} & \colhead{$\beta$} & \colhead{$\delta^{-1}$} & \colhead{$r$} & \colhead{$\nu_{\rm b}$} & \colhead{$\kappa=\frac{T_{\rm off}}{\tau}$} & \colhead{$\alpha_{\rm inj}$} & \colhead{$\chi^2$/DOF} & \colhead{$\tau$} & \colhead{$B_{\rm ADAF}$} \\
\colhead{} & \colhead{} & \colhead{$\SI{e9}{M_\odot}$} & \colhead{$\SI{e-2}{\frac{M_\odot}{yr}}$} & \colhead{} & \colhead{} & \colhead{$R_S$} & \colhead{GHz} & \colhead{} & \colhead{} & \colhead{} & \colhead{kyr} & \colhead{G}\\
\colhead{(1)} & \colhead{(2)} & \colhead{(3)} & \colhead{(4)} & \colhead{(5)} & \colhead{(6)} & \colhead{(7)} & \colhead{(8)} & \colhead{(9)} & \colhead{(10)} & \colhead{(11)} & \colhead{(12)} & \colhead{(12)}}
\startdata
1 & CI & $2.17^{+0.17}_{-0.13}$ & $1.20^{+0.73}_{-0.33}$ & $0.50$ & $1832^{+1122}_{-490}$ & $3.0$ & $15.5^{+5.9}_{-5.1}$ & $0.97^{+0.03}_{-0.32}$ & $0.50$ & 9.3/9 & $1.2^{+0.2}_{-0.3}$ & $815^{+189}_{-112}$ \\
2 & CI & $0.28^{+0.32}_{-0.13}$ & $9.59^{+7.80}_{-5.10}$ & $0.50$ & $1741^{+910}_{-467}$ & $32.8^{+35.0}_{-19.2}$ & $14.8^{+6.0}_{-4.9}$ & $0.93^{+0.07}_{-0.28}$ & $0.50$ & 11.6/8 & $1.3^{+0.2}_{-0.3}$ & $339^{+165}_{-110}$ \\
3  & CI & $1.34^{+0.24}_{-0.17}$ & $0.02^{+0.07}_{-0.01}$ & $0.50$ & $37^{+103}_{-23}$ & $3.0$ & $15.6^{+6.1}_{-5.3}$ & $0.97^{+0.03}_{-0.34}$ & $0.50$ & 9.3/9 & $1.2^{+0.2}_{-0.3}$ & $151^{+122}_{-50}$ \\
4 & CI & $2.93^{+0.23}_{-0.21}$ & $37.63^{+14.68}_{-9.43}$ & $0.99$ & $1680^{+666}_{-402}$ & $3.0$ & $15.1^{+5.9}_{-5.0}$ & $0.93^{+0.07}_{-0.28}$ & $0.50$ & 9.7/9 & $1.3^{+0.2}_{-0.3}$ & $578^{+92}_{-68}$ \\
5 & JP & $2.23^{+0.16}_{-0.13}$ & $1.16^{+0.62}_{-0.31}$ & $0.50$ & $1821^{+976}_{-484}$ & $3.0$ & $20.1^{+5.3}_{-3.8}$ &  & $0.50$ & 8.9/10 & $1.1^{+0.1}_{-0.1}$ & $822^{+171}_{-111}$ \\
6 & KP & $2.23^{+0.16}_{-0.13}$ & $1.16^{+0.61}_{-0.31}$ & $0.50$ & $1830^{+959}_{-489}$ & $3.0$ & $12.5^{+3.4}_{-2.4}$ &  & $0.50$ & 9.4/10 & $0.6^{+0.1}_{-0.1}$ & $821^{+170}_{-110}$ \\
7 & CI & $2.26^{+0.29}_{-0.15}$ & $1.24^{+1.49}_{-0.38}$ & $0.50$ & $1962^{+2393}_{-584}$ & $3.0$ & $11.1^{+4.4}_{-2.6}$ & $0.50$ & $0.50^{+0.00}_{-0.00}$ & 9.7/10 & $1.5^{+0.2}_{-0.2}$ & $847^{+346}_{-130}$ \\
8 & CI & $2.18^{+0.16}_{-0.12}$ & $1.15^{+0.63}_{-0.30}$ & $0.50$ & $1837^{+990}_{-478}$ & $3.0$ & $5.1^{+2.1}_{-1.5}$ & $0.00$ & $0.50$ & 12.2/10 & $2.2^{+0.4}_{-0.4}$ & $829^{+178}_{-109}$ \\
9 & JP & $2.23^{+0.16}_{-0.13}$ & $1.15^{+0.61}_{-0.30}$ & $0.50$ & $1823^{+950}_{-472}$ & $3.0$ & $26.1^{+8.0}_{-5.8}$ &  & $0.55$ & 10.2/10 & $1.0^{+0.1}_{-0.1}$ & $821^{+170}_{-108}$ \\
10 & JP & $2.24^{+0.16}_{-0.13}$ & $1.15^{+0.59}_{-0.31}$ & $0.50$ & $1816^{+929}_{-480}$ & $3.0$ & $10.6^{+4.8}_{-2.5}$ &  & $0.31^{+0.11}_{-0.12}$ & 6.7/9 & $1.5^{+0.3}_{-0.2}$ & $819^{+166}_{-110}$ \\
11  & JP & $2.25^{+0.17}_{-0.13}$ & $1.15^{+0.64}_{-0.32}$ & $0.50$ & $1806^{+1010}_{-491}$ & $3.0$ & $40.6^{+3.2}_{-5.3}$ &  & $0.75$ & 17.2/10 & $0.8^{+0.0}_{-0.1}$ & $817^{+176}_{-113}$ \\
\enddata
\tablecomments{Column (1) gives a reference number for the fitting run, as referred to in the text. Column (2) names the jet emission model that is employed (acronyms see text). ADAF model parameters are listed in columns (3) to (7), the SMBH mass (3), the mass accretion rate (4), the gas to total pressure (5), the inverse of the heat energy distribution between electrons and ions (6), and the inner truncation radius (7). Columns (8) to (10) refer to the jet emission model, specifically the break frequency (8), the remnant fraction  which only applies to the CI case (9), and the injection spectral index (10). Column (11) states the $\chi^2$ and degrees of freedom. Parameters without uncertainties are fixed to their stated values.  Column (12) is the derived aging time of the jet plasma following Eq. \ref{eq:aging}. Column (13) is the magnetic field in the ADAF disk based on the best fit parameters. }
\end{deluxetable*}

\begin{figure*}[t]
    \centering
    \includegraphics[width=0.99\textwidth]{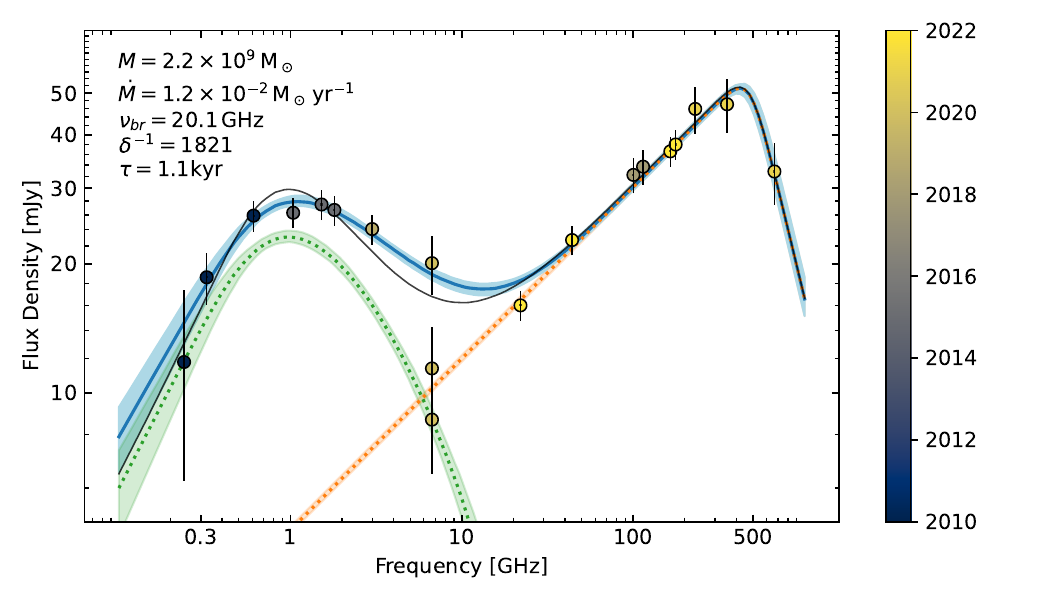}
    \caption{SED fit of our best fit JP Model (Run 5 in Tab. \ref{tab:sed_result}). The combined model is shown in blue, the self-absorbed jet model in green, and the ADAF model in orange. We show in black a CI model with a high duty cycle (Run 8). The two lower datapoints at 6.7\,GHz are the VLBA flux measurements of the compact and extended emission. The color scale of the datapoints shows the year when the observation was taken. }
    \label{fig:SED_JP}
\end{figure*}
We use the \verb|emcee| package (\citealp{Goodman2010-jw,Foreman-Mackey2013-hw}) to sample the posterior distribution of the MCMC. We run a total of 11 fits with different initial conditions, and list our results in Table \ref{tab:sed_result}. 
\begin{itemize}
    \item Run 1 is a Continuous Injection jet model with all parameters set to their default priors. In this case the we find a  mass $M$ that is only marginally larger than the one observed by \cite{Diniz2017-xw}, a relatively low accretion rate consistent with expectations and star formation rates, and break frequency between 10 to 20\,GHz. While the median of $\kappa$ is $0.78$, its distribution is highly asymmetric due to the parameter space boundary at 1. Therefore, we quote in all cases where $\kappa$ is free (Runs 1-4)  the most probable value for $\kappa$, which is very close to 1, implying the fit clearly favors a CI model that is off all the time (i.e., an AGN which has ceased to power its jets). 
    \item Run 2 is like 1 but has no prior on truncation radius $r$, which interestingly results in a much larger $r$, but also a mass $M$ that is close to the old value from \cite{David2009-hn}. The accretion rate in Run 2 is about 8 times larger than in 1, but closer to the star formation rate stated in \cite{Werner2014-vw}. 
    \item Run 3 is like 1 but has a uniform instead of normal prior on the ion/electron energy distribution parameter $\delta$. This causes $\delta$ to move to a value much closer to equal heat distribution between ions and electrons, and at the same time forces the accretion rate to be almost 2 orders of magnitude lower. 
    \item Run 4 is like 1 but has a minimal magnetic field pressure, meaning the magnetic field $B$ is 10 times lower. This increases the accretion rate by a factor of 30, which is unlikely since the star formation rate is very low (\citealp{Werner2014-vw}). 
    \item Run 5 is like 1 but uses a JP jet model instead of a CI model. Differences are small, since $\kappa$ in the CI model 1 is very close to $\kappa=1$ anyway. The break frequency is slightly larger, mostly due to the fact that the AGN is now off 100\% of the time. Despite having fewer free parameters (no $\kappa$) the $\chi^2$ is lowest (with the exception of the ``unphysical'' Run 10). We show this model in Fig. \ref{fig:SED_JP}.
    \item Run 6 is like 1 but uses a KP jet model instead of a CI model. This case is very similar to the JP model from Run 5, but has a significantly lower break frequency. However, the $\chi^2$ is slightly higher, and we favor the JP model over the KP. 
    \item Runs 7/8 are like 1 but have a fixed $\kappa$ to 0.5 / 0, meaning the AGN is active powering the jets half the time/all the time. While this is similar to the model explored by \cite{Schellenberger2020-ji}, our additional data, especially the VLA K and Q band measurements, slightly disfavor these models. The lower break frequency tries to compensate for the more active state of the AGN. 
    \item Runs 9/10/11 are like the standard JP case (Run 5) but with a different injection spectral index, $\alpha_{\rm inj}=0.55$ (Run 9), $\alpha_{\rm inj}=0.75$ (Run 11), and $\alpha_{\rm inj}$ free to vary (Run 10). Again, the data disfavors a steeper spectral index, and the fit tries to compensate through a much higher break frequency. In Run 10 however, the best fit injection index is around 0.3, which is shallower than expected. While this marks the lowest $\chi^2$, we do not think that an injection index of 0.3 is physical, and the fit is likely driven by the lower value of the VLA K band observation. 
\end{itemize}

\section{Discussion and outlook}\label{ch:discussion}

The information provided by the SED models allows us to further extract intrinsic properties of the central region near the AGN that are linked to the feeding and feedback processes. 
Despite the recently launched, small scale jets discovered by \cite{Schellenberger2020-ji} our SED fitting does not confirm that the AGN is actively powering jets at the moment. 
A continuous injection of particles into the jets, especially at a high duty cycle (small $\kappa$) appears less likely (see Tab. \ref{tab:sed_result}, Run 8, and Fig. \ref{fig:SED_JP} black line). 
The jet emission is consistent with an aging plasma as frequently observed in radio lobes. 
We can adopt the estimate for the magnetic field strength in the jets from \cite{Schellenberger2020-ji} of $B=\SI{4.7}{mG}$, which was derived assuming energy equipartition (\citealp{Govoni2004-dx,Giacintucci2008-wf}). The magnetic field strengths allows us compute the aging time  (see \citealp{Turner2018-qj}),
\begin{equation}\label{eq:aging}
    \tau = \frac{v B^{0.5}}{B^2 + B_{\rm IC}^2} \left[ \nu_b (1+z) \right]^{-0.5}~,
\end{equation}
where $\nu_b$ is the break frequency, $B_{\rm IC} = 0.317(1+z)^2\,\si{nT}$, and the constant $v = \left( \frac{243 \pi m_{\rm e}^5 c^2}{4 \mu_0^2 e^7} \right)^{0.5}$. 
For Run 5 we derive a timescale $\tau = \SI{1.1(1)}{kyr}$. 
The break frequency is somewhat correlated with the injection spectral index, and therefore we get the most extreme values for the timescale in Run 10 and 11, where the injection index is either 0.3 ($\tau = \SI{1.5(3)}{kyr}$) or 0.75 ($\tau = \SI{0.8(1)}{kyr}$). 
Therefore, the data slightly favor the hypothesis that  the AGN has stopped injecting electrons into the jets. 
The fact that this timescale is very different from the ADAF variability (1000 years vs. months) illustrates that the processes that launch jets are unrelated to the ``smaller'' feeding events of the black hole from surrounding cold gas (e.g., \citealp{Schellenberger2020-vl}). 
While in principle one could determine the magnetic field from the SSA turnover around 5\,GHz (see e.g., \citealp{Marscher1983-hg,Kim2021-lp}), uncertainties on the size of the absorbed core region dominate the results, which renders it impossible to even give an order of magnitude estimate (the size enters with the \nth{4} power). However, we believe the magnetic field on the order of mG is a realistic estimate of the average magnetic field along the $\sim$parsec jets (\citealp{O-Sullivan2010-en,Kim2021-lp}). 

VLBA data has shown that a compact core with an inverted spectrum is located between the jets (\citealp{Schellenberger2020-ji}). Our ADAF model provides an accurate description of the SED up to sub-mm wavelengths. Since the inverted spectrum is due to synchrotron emission one can derive the magnetic field within the ADAF emission region (\citealp{Mahadevan1997-rx}). 
For the JP model in run 5 we find a magnetic field of $B_{\rm ADAF} = \SI{870}{G}$, about $\num{200000}$ times larger than in the jets. 
It has been pointed out before that magnetic fields in accretion disks have to be 100\,s of G to launch jets (see \citealp{Blandford1982-zn,Jafari2019-yf}). 
For almost all Runs (Tab. \ref{tab:sed_result}) we find magnetic fields strengths in the ADAF emission region of 800 to $\SI{900}{G}$, with the exception of Run 2, 3 and 4: Run 2 has an almost 10 times smaller SMBH mass, and a much larger disk truncation, and if the magnetic field increases towards the center of the AGN, the average field will be smaller, in this case $339^{+165}_{-110}\,\si{G}$. Run 3 has an extremely low accretion rate, which implies the low magnetic field of $151^{+122}_{-50}\,\si{G}$. Run 4 has a high accretion rate (about 2000 times higher than run 3) but the magnetic pressure was set to be only 1\%, and the two effects almost balance each other, leading to a magnetic field of $578^{+92}_{-68}\,\si{G}$. 

If we want to precisely quantify the truncation radius $r$, which is tightly linked to other parameters in our fit, such as the SMBH mass, we need to have accurate measurements at even shorter wavelengths. It turns out that at wavelengths shorter than $\SI{450}{\um}$ (JCMT SCUBA2) the difference between models of Run 1 and 2 becomes very clear: While at lower frequencies the difference is minimal, at $\SI{1}{THz}$ or $\SI{300}{\um}$ the larger truncation radius (Run 2, with the smaller SMBH mass) has a 50\% larger flux than Run 1 (16.6 vs. $\SI{25.2}{mJy}$). This regime was accessible with Herschel. 

The lightcurve variability in NGC~5044 was reported previously, but our SMA observations have now quantified the level of variability. A 22\% overall scatter with a baseline around $\SI{46}{mJy}$ is significantly larger than the typical fractional uncertainty of the measurements of $\lesssim 2\%$. 
At the beginning of our program in 2021 we observed a $\sim 3$ month peak in the lightcurve, characterized by an increase from 45 to $\SI{55}{mJy}$. If we assume that changes in the accretion rate are the sole cause for this increase, it means that the ADAF accretion rate rose by 50\% to $\SI{1.74e-2}{M_\odot\,yr^{-1}}$. 
Our DRW model predicts a characteristic timescale of 67 days, similar to what was observed during the first peak.  
We can convert the typical variability timescale into a physical length, which yields $\sim 350\,R_{\rm S}$ ($\SI{0.08}{pc}$ or $\SI{0.5}{mas}$). This size is reasonable for the ADAF emitting accretion disk, and it is possible to resolve these scales with VLBA imaging at K/Q band (Schellenberger et al. in prep). 

We have demonstrated that the variability is real and has a characteristic timescale between 60 and 140 days, which is not explained by just random noise combined with our sparse sampling of the lightcurve. 
While the Arevalo test in section \ref{ch:periodogram} reveals a periodicity, the DRW model shows that the data are consistent with a random walk, which is not a periodic process. 
Results on the existence of Quasi-periodic oscillations (QPO) in X-ray lightcurves of AGN have been controversial and a random walk noise (Brownian noise) as an alternative interpretation appears more likely (e.g., \citealp{Press1978-du, MacLeod2010-ly, Kelly2009-mi, Krishnan2021-my,Rueda2022-ze}). 
In order to clearly characterize a truly periodic process that causes the observed variability, we will need a larger dataset with a similar sampling over several more years. 

While an ADAF-like spectrum has been observed in several sources, group central galaxies with their strong cooling and shallower gravitational potential than large clusters, mark a special class. 
Although the mm emission in the central AGN in NGC~5044 is well described by an ADAF model, the suitability has to be demonstrated for more cases. If the ADAF model turns out to be a common paradigm for the cooling and AGN feeding process, the overall feedback model can be linked from cold, infalling material, to large scale outflows and cavities (e.g., \citealp{Morganti2017-sa,Eckert2021-mm}). 
Our ongoing SMA program to search for other ADAFs in group central galaxies  starts from a small sample of selected galaxy groups with a wide range of properties (cool-core, relaxed, disturbed, CO rich), and has already confirmed an ADAF spectrum in NGC~5084 (Schellenberger et al., in prep.). This X-ray faint but high richness group hosts an S0 central galaxy with a radio point source and a large HI disk. However, no CO is detected (see \citealp{Kolokythas2022-te, OSullivan2018-yk}). In many aspects NGC~5084 is very different from NGC~5044. Further observations leading to potentially more detections in other groups will allow linking this radiatively inefficient accretion model to the cooling gas and the broader feedback process.

\section{Summary}\label{ch:summary}
We presented results from our recent SMA monitoring observations of the central continuum source in NGC~5044, and our VLA K and Q band, as well as archival JCMT observations. 
With an angular distance of only 31\,Mpc this system is ideal to study AGN feedback on the galaxy group scale. 
The mm-wavelength time variability has been proposed in the past, but it has never been studied in a systematic way so it can be linked to the feeding process of the AGN. Based on past SED fits this system was known host an AGN in ADAF mode, but with our new data we were able to quantify important parameters to better precision than previously. Our findings are as follows.

\begin{itemize}
    \item Our new SMA data undoubtedly demonstrates the mm-variability of the NGC~5044 continuum source over a 3 year timescale with measurements on monthly cadence whenever the source was observable.  We find a baseline flux of $\SI{46}{mJy}$ and an overall 22\% variability. We were able to identify distinct peaks in the lightcurve, such as the 20\% increase in early 2021. 
    \item We provide a statistical analysis of the lightcurve variability that takes into account the sparse and unequal sampling, and find consistency with a damped random walk with a characteristic timescale of 67 days, or a periodic signal of 136 days. Both timescales are consistent within their uncertainties.
    \item Based on the lightcurve we ``correct'' fluxes from other instruments for the SED fitting, which is well represented by a jet model (JP with SSA) and an AGN ADAF model. We are able to constrain the break frequency of the jet emission, which implies that the jet plasma is about $\SI{1}{kyr}$ old, and likely no longer powered by the AGN.  
    The SMBH mass derived from the ADAF model is consistent with recent estimates, and the accretion rate is in line with expectations. The magnetic field pressure in the ADAF emission region is significant, on the same order of the gas pressure. 
    \item With a few assumptions we are able to quantify the magnetic field in the jets ($\SI{4.7}{mG}$) and the accretion disk ($\sim \SI{870}{G}$). Based on the observed variability we can estimate the ADAF disk diameter to about $\SI{0.08}{pc}$ or $\SI{0.5}{mas}$, which can be resolved by high frequency VLBA observations. 
\end{itemize}
The rich multiwavelength dataset available for NGC~5044 makes this source unique, and allows us to push our understanding of the accretion and feedback process in galaxy groups. 
Our analysis and results show how complex AGN/feedback processes are, with a broader parameter space to be explored, i.e. longer lightcurve sampling, observations at higher frequencies, and a thorough comparison to other detected ADAFs in central group galaxies.

\begin{acknowledgments}
We acknowledge the Smithsonian Combined Support for Life on a Sustainable Planet, Science, and Research administered by the Office of the Under Secretary for Science and Research. 
GS acknowledges support from the Chandra High Resolution Camera Project through NASA contract NAS8-03060. 
Basic research in radio astronomy at the Naval Research Laboratory is supported by 6.1 Base funding.

The Submillimeter Array is a joint project between the Smithsonian Astrophysical Observatory and the Academia Sinica Institute of Astronomy and Astrophysics and is funded by the Smithsonian Institution and the Academia Sinica. We recognize that Maunakea is a culturally important site for the indigenous Hawaiian people; we are privileged to study the cosmos from its summit. We thank the staff of the SMA that made these observations possible.
The National Radio Astronomy Observatory is a facility of the National Science Foundation operated under cooperative agreement by Associated Universities, Inc.
The James Clerk Maxwell Telescope is operated by the East Asian Observatory on behalf of The National Astronomical Observatory of Japan; Academia Sinica Institute of Astronomy and Astrophysics; the Korea Astronomy and Space Science Institute; the National Astronomical Research Institute of Thailand; Center for Astronomical Mega-Science (as well as the National Key R\&D Program of China with No. 2017YFA0402700). Additional funding support is provided by the Science and Technology Facilities Council of the United Kingdom and participating universities and organizations in the United Kingdom and Canada.
Additional funds for the construction of SCUBA-2 were provided by the Canada Foundation for Innovation.
\end{acknowledgments}
\facility{SMA, VLA, JCMT}
\bibliography{paperpile_remote}
\bibliographystyle{aasjournal}

\appendix
\setcounter{table}{0}
\renewcommand{\thetable}{A\arabic{table}}
\section{Lomb Scargle periodogram}\label{ch:ls}
\begin{figure}[t]
    \centering
    \includegraphics[width=0.49\textwidth]{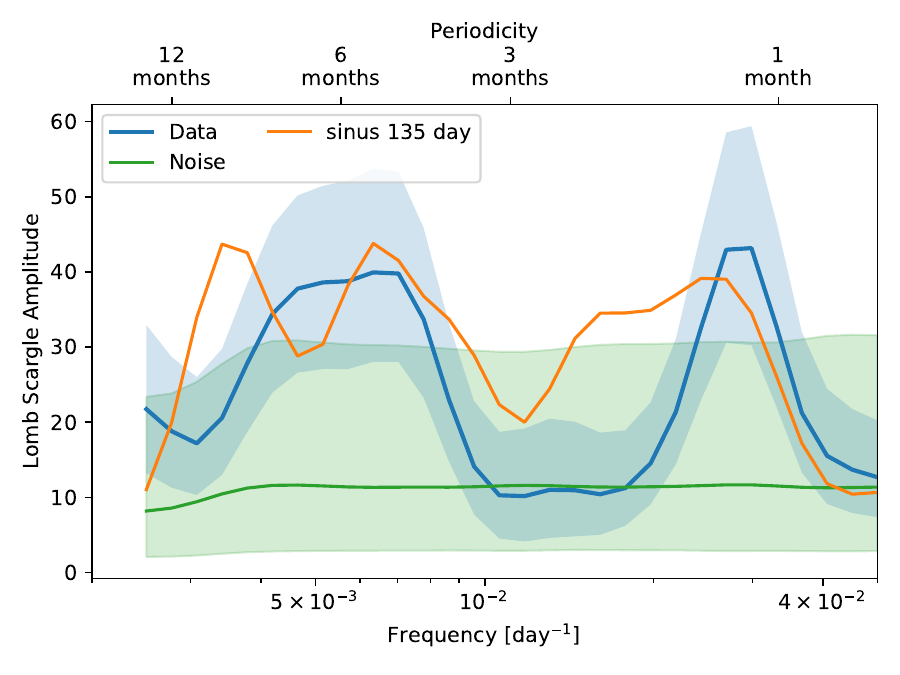}
    \caption{Lomb Scargle periodogram of the SMA lightcurve (blue). The green area shows the 1$\sigma$ random noise region, and the orange line the expected signal from a sine with 135 day period and the same observing mask as the SMA data.}
    \label{fig:LS}
\end{figure}
The Lomb-Scargle periodogram, initially based on \cite{Lomb1976-mb,Scargle1982-gc}, and with modifications by \cite{Press1989-so,Zechmeister2009-gz,Townsend2010-un,VanderPlas2018-hj}, is a standard tool for lightcurve analyses. We use the \verb|scipy| implementation and apply the mean-subtraction (``precenter''). We note that, while the input requires angular frequencies ($\omega = \frac{2\pi}{T}$, where $T$ is the period of a periodic signal) we refer to normal frequency, $f = \frac{1}{T}$, in our interpretation and plots. The Lomb-Scargle periodogram was specifically designed to detect periodic signals in unevenly spaced observations.
We utilize this tool and provide our SMA measurements as input. Figure \ref{fig:LS} (blue line) shows the output periodogram when sampling frequencies between one week and 400 days: We visually identify two distinct peaks, one at around 6 months, and a second at 1.5 months. 
To understand the meaning of these peaks we create simulated signals based on a) a flat lightcurve (47\,mJy) with a 22\% Gaussian noise (observed scatter), and b) a superposition of a flat lightcurve ($\SI{47}{mJy}$), plus a sine signal with a 135 day period. We sample these models on a daily base for a 4 year timerange (January 2021 to January 2025) and use the actual 30 SMA observing days as a mask. We compute 1000 Lomb-Scargle periodograms from random realizations and analyze the 68\% region around the median. For a) we find no periodicity and a flat signal (see green bar in Fig. \ref{fig:LS}), while for b) we find several peaks, two of which resemble the observations quite well (orange line in Fig. \ref{fig:LS}). 
This shows that due to the sparse lightcurve, it is possible that a periodic signal is causing the peaks that are observed, but it could also be explained by random noise since the significance of the blue line in Fig. \ref{fig:LS} is small. 

\section{Tables}
In this section we list all supplemental information, such as observation times and setups. Table \ref{tab:SMA} lists the SMA observations, and Tab. \ref{tab:smaflux} the corresponding flux densities. Table \ref{tab:jcmt} shows the JCMT observations.  
\begin{deluxetable*}{cccccccc}
 \tablecaption{SMA monitoring observations of NGC~5044\label{tab:SMA}}

\tablehead{\colhead{SMA} & \colhead{Date} & \colhead{Observation} & \colhead{On-source} & \colhead{\#Ant.} & \colhead{Flux Cal} & \colhead{Band Cal} & \colhead{Receiver}\\
\colhead{project} & \colhead{M/D/Y} & \colhead{[h:mm]} & \colhead{[h:mm]} & \colhead{} & \colhead{} & \colhead{} & \colhead{[GHz]}\\
\colhead{(1)} & \colhead{(2)} & \colhead{(3)} & \colhead{(4)} & \colhead{(5)} & \colhead{(6)} & \colhead{(7)} & \colhead{(8)}}
\startdata
2020B-S006 & 01/15/2021 & 4:00 & 2:26 & 7 & Vesta & 3C279 & 211--221, 231--241 \\
 & 02/07/2021 & 3:27 & 2:41 & 7 & Vesta & 3C279 & 207--217, 221--251 \\
 & 02/09/2021 & 6:15 & 2:31 & 7 & Vesta & 3C279 & 211--221, 231--241 \\
 & 03/05/2021 & 3:47 & 2:41 & 7 & Vesta & 3C279 & 210--221, 230--241 \\
 & 04/17/2021 & 2:36 & 1:28 & 6 & Vesta & 3C279 & 205--215, 225--235 \\
 & 04/30/2021 & 4:03 & 2:55 & 6 & Vesta & 3C279 & 211--229, 231--249 \\
 & 05/25/2021 & 8:31 & 4:52 & 5 & Vesta & 3C279 & 210--220, 230--240 \\
 & 05/31/2021 & 2:28 & 1:44 & 6 & Vesta & 3C279 & 211--229, 231--249 \\
 &  &  &  &  &  &  \\
2021A-S037 & 06/16/2021 & 3:48 & 2:13 & 6 & Vesta & 3C279 & 225--267 \\
 & 07/07/2021 & 3:12 & 2:12 & 6 & Vesta & 3C279 & 220--230, 240--250 \\
 & 07/19/2021 & 3:51 & 2:41 & 6 & Vesta & 3C279 & 211--221, 231--241 \\
 &  &  &  &  &  &  \\
2021B-S007 & 01/11/2022 & 5:24 & 3:25 & 6 & Ceres & 3C279 & 211--221, 231--241 \\
 & 01/27/2022 & 4:48 & 3:25 & 6 & Vesta & 1743-038 & 211--221, 231--241 \\
 & 03/06/2022 & 4:08 & 3:14 & 5 & Vesta & 1159+292 & 211--221, 231--241 \\
 & 04/11/2022 & 3:19 & 1:57 & 6 & Ceres & 3C279 & 220--230, 240--250 \\
 & 06/03/2022 & 2:40 & 0:44 & 6 & Ceres & 3C279 & 210--220, 230--240 \\
 &  &  &  &  &  &  \\
2022A-S004 & 06/24/2022 & 4:25 & 2:55 & 6 & Titan & BL Lac & 210--220, 230--240 \\
 &  &  &  &  &  &  \\
2022B-S003 & 01/11/2023 & 7:20 & 4:23 & 6 & Ceres & 3C279 & 211--221, 231--241 \\
 & 02/15/2023 & 5:53 & 3:24 & 5 & Ceres & 3C279 & 211--221, 231--241 \\
 & 03/22/2023 & 6:30 & 4:24 & 6 & Ceres & 3C279 & 211--221, 231--241 \\
 & 04/01/2023 & 5:57 & 4:07 & 6 & Ceres & 3C279 & 205--215, 225--235 \\
 & 04/08/2023 & 9:39 & 6:34 & 6 & Ceres & 3C279 & 210--221, 230--241 \\
 & 04/30/2023 & 9:26 & 6:34 & 6 & Ceres & 3C279 & 211--221, 231--241 \\
 & 05/24/2023 & 9:00 & 4:52 & 5 & Ceres & 3C279 & 211--221, 231--241 \\
 &  &  &  &  &  &  \\
2023B-S008 & 12/26/2023 & 4:19 & 2:49 & 6 & Pallas & 3C279 & 210--251 \\
 & 01/26/2024 & 5:43 & 3:54 & 7 & Ceres & 3C279 & 211--221, 231--241 \\
 & 02/27/2024 & 2:19 & 1:32 & 6 & Ceres & 3C84 & 210--220, 230--275 \\
 & 03/26/2024 & 6:47 & 3:11 & 7 & Pallas & 3C279 & 211--221, 231--241 \\
 & 04/23/2024 & 4:16 & 1:44 & 5 & Ceres & BL Lac & 210--220, 230--240 \\
 & 05/06/2024 & 3:48 & 2:04 & 7 & Vesta & 3C279 & 211--221, 231--241
\enddata
\tablecomments{Column (1) is the SMA project/proposal numbers and semesters,  column (2) lists the observing date, columns (3) and (4) state the total and on-target observing time, respectively. The active number of antenna during the observation is listed in column (5), and the flux and bandpass calibration sources are given in columns (6) and (7), respectively. Column (8) gives the frequency of the sidebands. }
\end{deluxetable*}

\begin{deluxetable*}{ccc|cccc|cccc}
 \tablecaption{SMA 230\,GHz fluxes of NGC~5044\label{tab:smaflux}}
 \tablehead{
 \colhead{Date} &
 \colhead{Flux} &
 \colhead{Flux Error} &
 \colhead{} &  
 \colhead{Date} &
 \colhead{Flux} &
 \colhead{Flux Error} &
 \colhead{} &  
 \colhead{Date} &
 \colhead{Flux} &
 \colhead{Flux Error}  \\
 \colhead{Y-M-D}&
 \colhead{mJy} & 
 \colhead{mJy} & 
 \colhead{} & 
 \colhead{Y-M-D}&
 \colhead{mJy} & 
 \colhead{mJy} & 
 \colhead{} & 
 \colhead{Y-M-D}&
 \colhead{mJy} & 
 \colhead{mJy}   \\
 \colhead{(1)}&
 \colhead{(2)} & 
 \colhead{(3)} & 
 \colhead{} & 
 \colhead{(1)}&
 \colhead{(2)} & 
 \colhead{(3)} & 
 \colhead{} & 
 \colhead{(1)}&
 \colhead{(2)} & 
 \colhead{(3)} 
 }
\startdata
2021-01-15 & 46.15 & 0.58 & & 2021-07-19 & 38.72 & 2.23 & & 2023-04-01 & 38.72 & 1.61 \\
2021-02-07 & 47.94 & 0.58 & & 2022-01-11 & 47.77 & 0.58 & & 2023-04-08 & 47.77 & 1.39 \\
2021-02-09 & 45.21 & 0.64 & & 2022-01-27 & 43.85 & 0.51 & & 2023-04-30 & 43.85 & 1.11 \\
2021-03-05 & 45.29 & 0.95 & & 2022-03-06 & 44.02 & 0.49 & & 2023-05-24 & 44.02 & 1.95 \\
2021-04-17 & 52.21 & 0.64 & & 2022-04-11 & 43.28 & 2.15 & & 2023-12-26 & 43.28 & 0.60 \\
2021-04-30 & 55.52 & 1.23 & & 2022-06-03 & 51.84 & 3.04 & & 2024-01-26 & 51.84 & 1.50 \\
2021-05-25 & 51.49 & 2.89 & & 2022-06-24 & 44.41 & 0.59 & & 2024-02-27 & 44.41 & 0.75 \\
2021-05-31 & 54.11 & 1.07 & & 2023-01-11 & 42.24 & 1.09 & & 2024-03-26 & 42.24 & 0.97 \\
2021-06-16 & 45.10 & 5.12 & & 2023-02-15 & 45.42 & 0.78 & & 2024-04-23 & 45.42 & 1.09 \\
2021-07-07 & 48.26 & 0.97 & & 2023-03-22 & 46.86 & 0.51 & & 2024-05-06 & 46.86 & 1.86 \\
\enddata
\tablecomments{For each observing date (1) we list the measured SMA fluxes (2) and uncertainties (3).}
\end{deluxetable*}

\begin{deluxetable*}{ccccc|cccc}
 \tablecaption{Archival JCMT Observations of NGC~5044\label{tab:jcmt}}
 \tablehead{
 \colhead{Project} &
 \colhead{Date} &
 \colhead{Exposure}&
 \colhead{Flux 850$\si{\mu m}$} &
 \colhead{} &
 \colhead{Project} &
 \colhead{Date} &
 \colhead{Exposure}&
 \colhead{Flux 850$\si{\mu m}$} \\
 \colhead{} & 
 \colhead{} & 
 \colhead{850/450$\si{\mu m}$}& 
 \colhead{mJy} &
 \colhead{} &
 \colhead{} & 
 \colhead{} & 
 \colhead{850/450$\si{\mu m}$}& 
 \colhead{mJy}\\
 \colhead{(1)} & 
 \colhead{(2)} & 
 \colhead{(3)}& 
 \colhead{(4)} &
 \colhead{} &
 \colhead{(1)} & 
 \colhead{(2)} & 
 \colhead{(3)}& 
 \colhead{(4)}
 }
\startdata
S14AU03 & 2014-06-08 & 1.1/0.3min&$\num{66.1(37)}$ & & M16AP083 & 2016-02-02 & 2.9/0.7min&$\num{59.6(33)}$\\ 
S14BU03 & 2015-01-25 & 1.1/0.3min&$\num{85.5(37)}$ & &        & 2016-02-17 & 2.8/0.7min&$\num{58.9(33)}$\\
M15AI70 & 2015-04-27 & 1.2/0.3min&$\num{67.7(26)}$ & &        & 2016-04-17 & 2.8/0.7min&$\num{56.2(33)}$\\
        & 2015-06-17 & 1.5/0.4min&$\num{45.2(31)}$ & &         & 2016-04-28 & 2.8/0.7min&$\num{59.2(33)}$\\
M15BI025 & 2015-12-17 & 2.9/0.7min&$\num{57.5(33)}$ & &M19AP054 & 2019-04-26 & 2.9/0.7min&$\num{59.5(34)}$\\
         & 2015-12-25 & 2.9/0.7min&$\num{46.8(31)}$ & &M19BP038 & 2019-12-19 & 2.8/0.7min&$\num{61.3(34)}$\\
         & 2016-01-12 & 2.9/0.7min&$\num{55.5(32)}$ & & M20AP043 & 2020-05-22 & 2.9/0.7min&$\num{60.4(34)}$\\
& & & & &  E21AK006 & 2021-02-02 & 2.9/0.7min&$\num{47.1(31)}$ 
\enddata
\tablecomments{We list the JCMT projects (1), the observation date (2), effective exposure times for the two frequencies (3), and the flux measurement including the uncertainty (4).}
\end{deluxetable*}

\end{document}